\documentclass[11pt,a4paper]{article}

\usepackage[utf8]{inputenc}
\usepackage[T1]{fontenc}
\usepackage{amsmath,amssymb}
\usepackage{graphicx}
\usepackage{booktabs}
\usepackage{hyperref}
\usepackage{xcolor}
\usepackage{natbib}
\usepackage[margin=1in]{geometry}
\usepackage{caption}
\usepackage{subcaption}
\usepackage{float}
\usepackage{multirow}
\usepackage{array}

\hypersetup{colorlinks=true,linkcolor=blue,citecolor=blue,urlcolor=blue}

\title{A Diagnostic Evaluation Framework for AI-Generated Cover Songs\\Using Music-Theoretic and Acoustic Features}

\author{Yingxin Liang\\
Hangzhou Xiaoying Innovation Technology Co., Ltd.\ (Rythmix AI)\\
\texttt{yingxinliang.yxl@gmail.com}}

\date{}

\begin{document}
\maketitle

\begin{abstract}
AI-generated covers often fail through local musical errors that a global quality score cannot locate: the vocal contour may remain recognizable while the accompaniment uses the wrong harmonic function, or the output may stay in key while the arrangement remains incomplete. We present a five-dimensional diagnostic framework covering melodic pitch, harmonic progression, key consistency, style consistency, and arrangement/production quality. The benchmark contains 30 covers generated from 5 source songs by 6 systems, with expert severity ratings and 9 symbolic or acoustic features. Harmonic progression and arrangement had the highest severe-error rates (53\% and 47\%), whereas key consistency was better preserved. Six covers combined acceptable key consistency with severe harmonic errors. Large-leap ratio had a nominal association with melodic ratings (Spearman $\rho = -0.429$, uncorrected $p = 0.018$), but no feature correlation survived the nine-test multiplicity reference. An interpretable percentile-rule pilot likewise failed to outperform a fixed majority baseline reliably across 16 dimension-level comparisons. The results separate useful diagnostic evidence from dependable automatic scoring: low-level and symbolic summaries can expose particular symptoms, but they do not replace context-aware musical judgment.
\end{abstract}

\section{Introduction}

AI-generated cover songs are becoming increasingly capable of producing recognizable renditions of existing songs with changed timbre, genre, or production style. However, the resulting outputs often fail in ways that are difficult to diagnose with a single global quality score. A cover may follow the original vocal melody while using an implausible chord progression; it may remain broadly within the correct key while failing to realize the intended style; or it may sound loudness-normalized while still having an empty, unbalanced, or musically incomplete arrangement. In practice, this can produce a cover that does not sound simply ``out of key''; the more specific problem is that the accompaniment does not know where the phrase is going. These are distinct musical failure modes, and collapsing them into one perceptual rating makes the resulting evaluation less useful for model analysis.

This problem is especially important for cover-song generation because the task is reference-based. Unlike open-ended text-to-song generation, cover generation involves an existing song whose melody, tonality, harmonic expectations, and structure can be used as diagnostic anchors. This makes it possible to evaluate not only whether an output sounds good, but also where and how it deviates from the source or from the target style.

Existing evaluation benchmarks for AI-generated music primarily focus on perception-level assessment of text-to-song outputs. SongBench \citep{wu2026songbench} provides multi-dimensional quality scoring over large-scale generated song samples, while SongEval \citep{lei2025songeval} evaluates aesthetic dimensions such as coherence, memorability, and musicality. These benchmarks are valuable for measuring overall listening quality, but they are not designed to localize music-theoretic errors in reference-based cover generation.

Recent generation work has begun to encode explicit music-theoretic constraints. Decomposed chord-generation approaches, for example, separate retrieval, editing, and reranking to improve harmonic feasibility. Yet evaluation still lacks a way to expose errors separately in pitch, harmony, key stability, style, and arrangement.

This paper addresses that gap through three contributions:
\begin{enumerate}
    \item We propose a five-dimensional diagnostic taxonomy for AI-generated cover-song evaluation, covering melodic pitch accuracy, harmonic progression validity, key consistency, style consistency, and arrangement/production quality.
    \item We construct a 30-sample diagnostic benchmark from 5 source songs and 6 generation systems, with severity-graded expert annotations and time-stamped error notes.
    \item We test nine automatic features and an interpretable LOO rule pilot. Large-leap ratio has a nominal association with melodic judgments, but the rule system does not beat a fixed majority baseline reliably; the D3-high/D2-low cases further show why tonal stability cannot stand in for harmonic correctness.
\end{enumerate}

\section{Related Work}

\paragraph{AI music evaluation benchmarks.}
Recent benchmarks have made AI-generated songs easier to compare at scale, especially for text-to-song generation. SongBench \citep{wu2026songbench} evaluates quality across dimensions including vocal quality, instrument quality, melody, structure, arrangement, mixing, and musicality. SongEval \citep{lei2025songeval} focuses on aesthetic dimensions such as coherence, memorability, breath naturalness, structural clarity, and overall musicality. Concurrent work extends single-point MOS prediction with stem-aware cross-attention and interval-based uncertainty modeling for full-song aesthetics \citep{lv2026songaesthetics}. The present study complements that learned scoring direction by testing where interpretable low-level and symbolic features fail to localize cover-specific musical errors. These approaches address listening-quality assessment at scale, but they do not use a reference song to ask a more local diagnostic question: which part of the cover failed, and relative to what musical expectation?

\paragraph{Music-theoretic constraints in generation.}
Harmonic feasibility has also become a generation-side concern. \citet{he2026chord}, for example, propose a retrieval-edit-rerank framework for chord generation that separates stylistic diversity from harmonic correctness. This is closely related to the present study because a generated cover may satisfy broad stylistic cues while still making local harmonic decisions that contradict the melody or phrase function.

\paragraph{Cover song generation.}
Cover generation sits between preservation and transformation: the source melody, phrase structure, and tonal expectations should remain recognizable, while voice identity, genre, instrumentation, or arrangement style may change. SongEcho \citep{li2026songecho} approaches cover-song generation through instance-adaptive element-wise linear modulation. ACE-Step and ACE-Step~1.5 \citep{gong2025acestep,gong2026acestep15} provide a broader music generation foundation model and conditioning extensions, including SA~ControlNet and MuseControlLite. The resulting outputs are therefore not well described by a single ``good/bad'' score; they require evaluation that can separate preserved elements from newly introduced errors.

\paragraph{Audio quality assessment and MIR tools.}
The MIR and audio-engineering toolchain provides useful evidence, but its scope is uneven. ITU-R BS.1770-4 defines integrated loudness measurement, while MUSHRA (ITU-R BS.1534) provides a subjective evaluation protocol for audio quality. Demucs \citep{defossez2021demucs}, Basic~Pitch \citep{bittner2022basicpitch}, pretty\_midi \citep{raffel2014prettymidi}, music21 \citep{cuthbert2010music21}, librosa \citep{mcfee2015librosa}, and pyloudnorm \citep{steinmetz2019pyloudnorm} make it possible to extract source-separated stems, symbolic pitch features, pitch-class evidence, loudness measures, and spectral descriptors. The open question is not whether these features are useful, but where they stop being sufficient: they can describe parts of the signal, but may not capture whether an arrangement actually works as music.

\section{Evaluation Framework}

\subsection{Diagnostic Dimensions}

The proposed framework evaluates AI-generated cover songs across five diagnostic dimensions. Each dimension corresponds to a different type of musical adequacy and is intended to localize the source of quality degradation.

\paragraph{D1: Melodic Pitch Accuracy.}
This dimension measures whether the generated vocal melody preserves the intended pitch contour and avoids salient wrong notes. Evidence includes vocal MIDI transcription, pitch-contour inspection, and large-leap ratio. Severe D1 failures include out-of-tune vocal lines, incorrect melodic intervals, octave displacement that changes melodic identity, or unstable pitch tracking that disrupts the recognizability of the source melody.

\paragraph{D2: Harmonic Progression Validity.}
This dimension measures whether the accompaniment follows a coherent harmonic progression and avoids non-functional or contextually inappropriate chord changes. Evidence includes chromagram visualization and time-stamped expert annotations identifying specific harmonic errors. Severe D2 failures include chords that contradict the expected function of a phrase, incorrect cadential harmony, or accompaniment that remains in a nominal key but fails to support the melody.

\paragraph{D3: Key Consistency.}
This dimension measures whether the generated cover maintains a stable or musically justified key center. Evidence includes key confidence, key-change rate, in-key note ratio, and chroma-based views. D3 differs from D2: a sample can stay in one key while still using poor chord progressions, and this distinction is central to the framework.

\paragraph{D4: Style Consistency.}
This dimension measures whether the generated cover matches the target style prompt in instrumentation, rhythm, harmonic language, texture, and overall musical character. Evidence comes from expert comparison between the intended style prompt and the generated output. No computational metric is assigned to D4 because genre and style matching require semantic understanding of musical conventions rather than only signal-level analysis.

\paragraph{D5: Arrangement and Production Quality.}
This dimension measures whether instrumentation, texture, balance, timbre, structure, and production quality support the intended cover. Evidence comes primarily from expert annotation, with loudness and spectral features reported as low-level reference indicators. Severe D5 failures include incomplete instrumentation, poor vocal--instrument balance, distorted timbre, empty texture, or structurally incoherent arrangement.

\subsection{Severity Grading}

Each dimension is rated on a 5-point scale. For analysis, the scores are grouped into three severity levels (Table~\ref{tab:severity}).

\begin{table}[h]
\centering
\caption{Severity grading scheme.}
\label{tab:severity}
\begin{tabular}{lcp{7cm}}
\toprule
Severity & Score & Meaning \\
\midrule
Acceptable & 4--5 & No salient error, or only minor issues that do not affect musical coherence \\
Minor issue & 3 & Perceptible issue, but the cover remains musically usable \\
Severe issue & 1--2 & Error affects musical coherence, recognizability, or user experience \\
\bottomrule
\end{tabular}
\end{table}

This severity grouping is designed to make the framework useful for both research analysis and practical model diagnosis. The exact numerical score captures relative quality, while the severity grouping identifies whether a failure would likely matter to listeners or downstream product use.

\subsection{Annotation Protocol}

The first author, who has absolute pitch, formal music-theory training, and professional experience evaluating AI-generated music systems at an AI music product company, performed all annotations. For each sample, the first author assigned five dimension scores and recorded time-stamped diagnostic notes describing the specific musical errors. The notes identify affected musical elements where applicable, including melodic deviations, chord errors, key changes, style mismatches, and arrangement or production defects.

To check rating stability, the first author re-annotated six samples (20\%) after a one-week interval. We treat this exercise as a procedural consistency check rather than a reliability estimate and do not report a formal agreement coefficient.

The use of a single expert annotator limits inter-rater reliability analysis, but the annotation protocol was designed to improve transparency: each low score must be supported by an explicit musical description and, where possible, a time-stamped location in the audio. This makes the benchmark auditable and provides a basis for future multi-annotator validation.

\section{Dataset}

\subsection{Source Songs}

Five source songs were selected to cover different musical conditions, including pop ballad, rock, electronic, and harmonically varied mixed-genre material. The selection criteria were: (a)~the evaluator could manually verify the source melody, chord progression, and key centers; (b)~the set included both relatively simple diatonic progressions and more complex harmonic materials; and (c)~the songs varied in tempo, vocal range, texture, and structural complexity. The purpose of the source-song set is not to represent all popular music, but to provide a controlled exploratory benchmark for testing whether the diagnostic framework can distinguish different failure modes.

\subsection{Generation Models}

Six generation systems were evaluated (Table~\ref{tab:models}). Each source song was processed by all 6 systems, producing 30 generated cover samples. All samples were generated with style prompts specifying target genre, instrumentation, and vocal characteristics. Because the systems differ in interface, controllability, and post-processing behavior, the benchmark is best interpreted as a diagnostic comparison of output behavior rather than a controlled architecture-level comparison.

\begin{table}[h]
\centering
\caption{Generation systems evaluated in this study.}
\label{tab:models}
\begin{tabular}{ll}
\toprule
Model & Description \\
\midrule
SongEcho & Instance-adaptive cover generation via ELM \\
ACE-Step+SA & ACE-Step 1.5 with SA ControlNet + LoRA \\
ACE-Step+MCL & ACE-Step 1.5 with MuseControlLite \\
ACE-Step & ACE-Step 1.5 base model \\
MiniMax & Commercial music generation API \\
Mureka & Commercial music generation API \\
\bottomrule
\end{tabular}
\end{table}

\subsection{Data Availability}

Due to copyright constraints on source songs and licensing terms of commercial APIs, the full audio set cannot be released without restrictions. We therefore release the annotation schema, extracted features, anonymized diagnostic metadata, and analysis pipeline. Audio excerpts are provided only where licensing permits.

\section{Computational Methods}

\subsection{Pipeline}

Each generated cover sample was processed through the following pipeline:

\begin{enumerate}
    \item \textbf{Source separation.} Demucs (\texttt{htdemucs\_ft}, two-stem mode) separates each cover into vocal and accompaniment stems.
    \item \textbf{Vocal transcription.} Basic Pitch transcribes the separated vocal stem into MIDI, producing note-level pitch, onset, and offset information. The resulting MIDI files are parsed with pretty\_midi for pitch-interval and range features, and with music21 for pitch-class and key-related analyses.
    \item \textbf{Feature extraction.} Nine features are computed from the separated stems, vocal MIDI transcription, and full audio mixture. librosa is used for spectral-contrast analysis, while pyloudnorm is used for loudness and loudness-range measurements.
\end{enumerate}

This pipeline is designed to provide diagnostic evidence rather than fully automatic scoring. The features are used to test whether low-level or symbolic measurements align with expert ratings.

\subsection{Feature Definitions}

\paragraph{D1-related features from vocal MIDI.}
\emph{Large-Leap Ratio (LLR):} Proportion of adjacent note intervals exceeding 7 semitones. Higher values indicate more melodically implausible jumps.
\emph{Pitch Range (PR):} Difference between the maximum and minimum transcribed vocal pitch, in semitones. PR is reported for reference and may be inflated by transcription artifacts.
\emph{Pitch Stability (PS):} Standard deviation of adjacent-note interval sizes. Higher values indicate more irregular intervallic behavior.

\paragraph{D2/D3-related features from vocal MIDI and chroma representations.}
\emph{In-Key Note Ratio (IKNR):} Proportion of transcribed vocal notes whose pitch classes belong to the detected global key's diatonic scale. The global key is estimated using the Krumhansl--Schmuckler key-finding algorithm implemented in music21.
\emph{Key Confidence (KC):} Correlation between the observed pitch-class distribution and the best-matching key template. Higher values indicate a more clearly defined tonal center.
\emph{Key-Change Rate (KCR):} Proportion of adjacent 10-second windows where the independently detected key changes. Higher values indicate less stable tonality.

\paragraph{D5-related features from the full mixture.}
\emph{Loudness Deviation (LUFS\_dev):} Absolute difference between integrated loudness and a reference level of $-12$~LUFS, computed using pyloudnorm following ITU-R BS.1770-4. The deviation is reported in LU.
\emph{Loudness Range (LRA):} Dynamic range in Loudness Units, computed from the stereo signal.
\emph{Spectral Contrast (SC):} Mean spectral contrast across all frequency bands and time frames, computed using librosa from a mono downmix.

\subsection{Excluded Features}

No automatic chord-recognition metric is used for D2. Available chord-recognition outputs were not sufficiently reliable for diagnostic evaluation of generated covers, especially when source separation artifacts, dense textures, or distorted timbres were present. A noisy chord-recognition metric would risk converting tool errors into false evidence about model behavior. D2 is therefore evaluated through time-stamped expert annotation, supported by chromagram inspection rather than automatic chord labels.

No computational metric is used for D4. Style consistency depends on genre conventions, instrumentation choices, rhythmic feel, and semantic alignment with the prompt. These properties are not adequately represented by the low-level features used in this study.

\subsection{Known Limitations of the Pipeline}

Basic Pitch transcription can introduce errors from instrumental bleed, octave confusion, or unstable vocal separation. These artifacts can affect PR, LLR, and PS. For this reason, the computational features are interpreted as supporting evidence rather than ground-truth measurements. Rank-order correlations are reported as exploratory statistics and should be validated on larger datasets with improved transcription and additional annotators.

\subsection{Rule-Based Diagnostic Pilot}

We tested whether the extracted features could support a transparent rule-based diagnostic rather than only correlational analysis. Expert ratings were collapsed into three severity classes: ratings 4--5 were coded as acceptable~(0), rating~3 as minor~(1), and ratings 1--2 as severe~(2). For each metric--dimension pair, two candidate cutpoints were estimated from the training data: a severity-2 quartile (the lower quartile for higher-is-worse metrics and the upper quartile for lower-is-worse metrics) and the severity-1 median. Small-sample overlap can reverse their raw order, so the pair was ordered by the known metric direction before assigning yellow and red decision regions. A programmatic assertion enforced yellow $<$ red for higher-is-worse metrics (LLR, PS, KCR, and LUFS\_dev) and yellow $>$ red for lower-is-worse metrics (IKNR, KC, LRA, and SC) in every validation fold. The archived JSON retains both raw candidates and ordered cutpoints (see Appendix~\ref{app:thresholds}).

Metrics were mapped to D1 (LLR and PS), D2 (IKNR as a deliberately coarse tonal proxy rather than a chord-function measure), D3 (IKNR, KC, and KCR), and D5 (LUFS\_dev, LRA, and SC). Each metric cast an acceptable, yellow, or red vote. A dimension was labeled severe when a strict majority of its metrics cast red votes, minor when a strict majority cast at least yellow votes, and acceptable otherwise. D4 was excluded because style consistency requires semantic comparison with the target prompt and is not represented by the available low-level or symbolic features.

Thresholds were evaluated by leave-one-out (LOO) validation: each fold learned cutpoints from 29 generated covers and applied them to the held-out cover. We evaluated the three-class task and a binary task that grouped acceptable and minor cases against severe cases. Accuracy and weighted $F_1$ were compared with a majority-class baseline. For bootstrap comparisons, the baseline was fixed to the full-sample majority class for each dimension and setting and was not recomputed within resamples. Defining the baseline on the full sample can slightly favor it; we retain that convention to match the reported baseline and state it explicitly.

Uncertainty was estimated with 10,000 paired generated-sample-level bootstrap draws and a fixed random seed. Each draw sampled 30 filenames with replacement, and the same indices were applied to rule and baseline predictions. Descriptive overall summaries carried all four diagnostic dimensions for each selected filename instead of resampling the 120 dimension rows independently. We report percentile 95\% intervals for $\Delta = \text{rule} - \text{baseline}$ and two-sided $p$-values computed as $2\min[P(\Delta > 0),\, P(\Delta < 0)]$. Weighted $F_1$ used \texttt{zero\_division=0} when a resample omitted a class.

\section{Results}

\subsection{Expert Rating Summary}

Table~\ref{tab:ratings} shows the mean expert ratings per model across the five diagnostic dimensions.

\begin{table}[t]
\centering
\caption{Mean expert ratings by model ($n = 5$ songs per model).}
\label{tab:ratings}
\begin{tabular}{lcccccc}
\toprule
Model & D1 & D2 & D3 & D4 & D5 & Mean \\
\midrule
MiniMax      & 4.8 & 4.0 & 4.8 & 3.2 & 3.6 & 4.1 \\
Mureka       & 4.2 & 3.8 & 4.6 & 4.6 & 3.6 & 4.2 \\
ACE-Step     & 2.8 & 2.0 & 4.2 & 4.2 & 2.6 & 3.2 \\
SongEcho     & 3.6 & 2.8 & 3.4 & 2.4 & 2.6 & 3.0 \\
ACE-Step+SA  & 2.6 & 1.4 & 3.0 & 1.6 & 1.6 & 2.0 \\
ACE-Step+MCL & 1.2 & 1.0 & 1.2 & 1.4 & 1.2 & 1.2 \\
\bottomrule
\end{tabular}
\end{table}

\begin{figure}[t]
\centering
\includegraphics[width=0.85\textwidth]{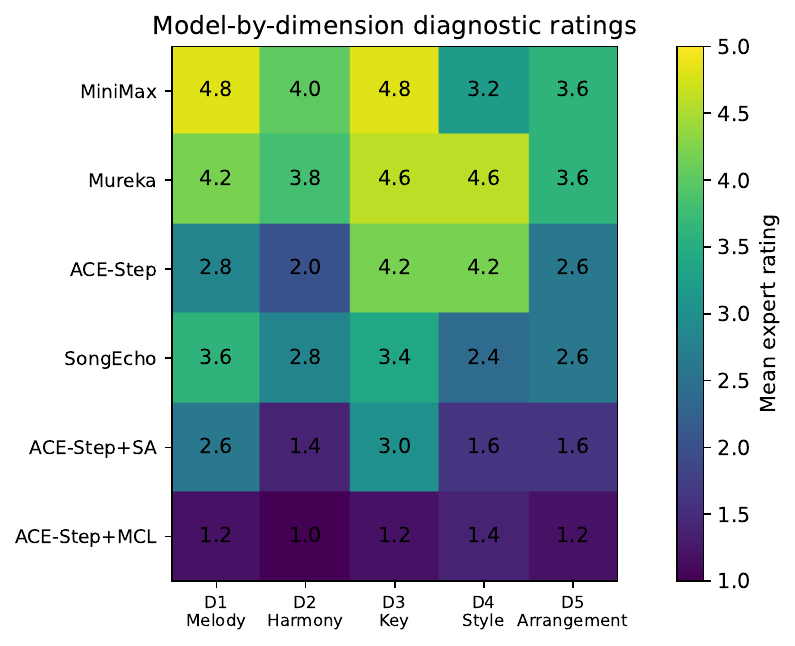}
\caption{Model-by-dimension heatmap. Each cell shows the mean expert rating for one model and one diagnostic dimension. MiniMax and Mureka retain relatively high ratings across most dimensions, ACE-Step base shows the D3-high/D2-low asymmetry at the model level, and ACE-Step+MCL fails consistently across dimensions.}
\label{fig:heatmap}
\end{figure}

MiniMax and Mureka achieved the highest overall ratings. ACE-Step+MCL produced severe failures in 3 of 5 songs, with all five dimensions scored~1 in those cases. ACE-Step base had relatively high key and style ratings (D3 = 4.2, D4 = 4.2) but much lower harmony and arrangement ratings (D2 = 2.0, D5 = 2.6). Broad tonal or stylistic plausibility therefore did not guarantee functional harmony or a complete arrangement.

\subsection{Severity Distribution}

Table~\ref{tab:severity_dist} shows the proportion of ratings at each severity level by dimension.

\begin{table}[h]
\centering
\caption{Severity distribution by dimension ($n = 30$ samples).}
\label{tab:severity_dist}
\begin{tabular}{lccc}
\toprule
Dimension & Acceptable (4--5) & Minor (3) & Severe (1--2) \\
\midrule
D1 Melody      & 47\% & 23\% & 30\% \\
D2 Harmony     & 23\% & 23\% & 53\% \\
D3 Key         & 63\% &  7\% & 30\% \\
D4 Style       & 37\% & 17\% & 47\% \\
D5 Arrangement & 23\% & 30\% & 47\% \\
\bottomrule
\end{tabular}
\end{table}

D2 had the highest severe-error rate (53\%), followed by D4 and D5 (47\% each). D3 was the best-preserved dimension, with 63\% of samples rated acceptable. This distribution indicates that the evaluated systems more often preserve a tonal center than produce coherent chord progressions, accurate style realization, or complete arrangements.

The score distribution had substantial mass at both endpoints: 26\% of the 150 dimension ratings were~1 and 22\% were~5. Moderate ratings were less dominant than clear successes or substantial failures.

\subsection{Inter-Dimension Correlations}

All ten pairwise inter-dimension Spearman correlations had uncorrected $p < 0.01$, ranging from $\rho = 0.507$ for D1--D4 to $\rho = 0.799$ for D3--D5 (see Table~\ref{tab:corr_matrix} in Appendix~B for the full matrix). We treat these comparisons as exploratory and make no familywise significance claim. The consistently positive associations show that failure modes often co-occur: covers with poor melody accuracy also tended to score poorly on harmony, style, or arrangement.

Co-occurrence does not make the dimensions interchangeable. Six of the 30 covers had D3 $\geq$ 4 but D2 $\leq$ 2: the tonal center remained plausible while the chord progression contained severe errors. Those cases show that a model can preserve surface-level tonality without generating functional harmony inside that tonal space.

\begin{figure}[t]
\centering
\includegraphics[width=0.7\textwidth]{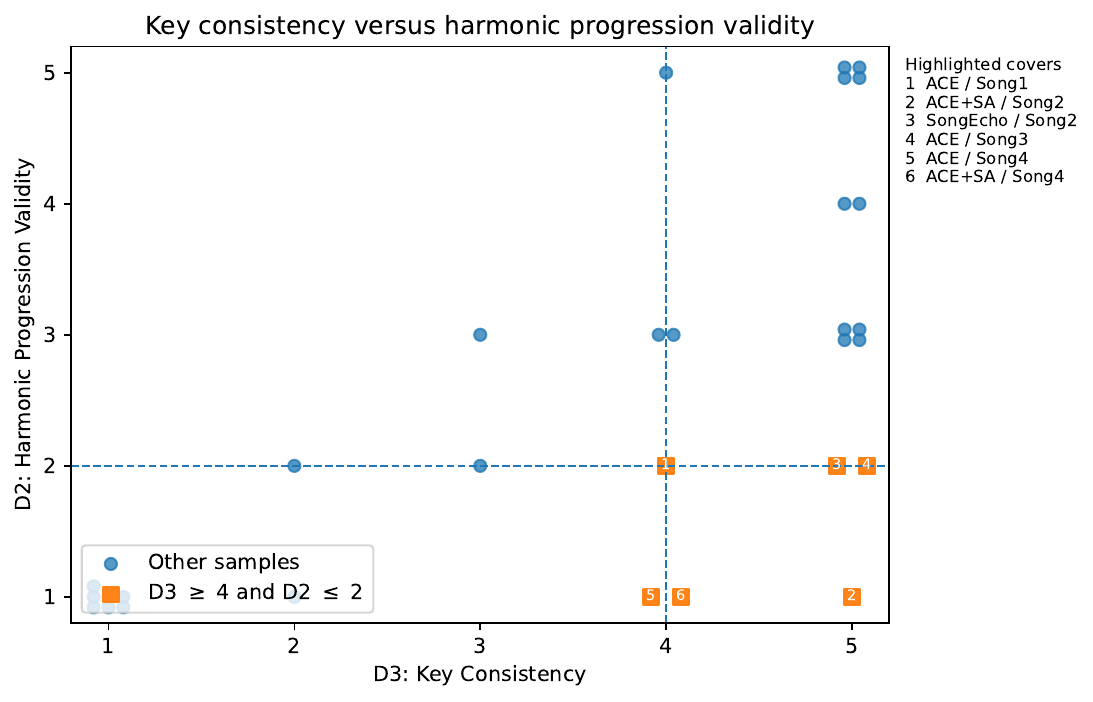}
\caption{D3 versus D2 sample-level scatter plot. Each point represents one generated cover sample. The lower-right diagnostic region corresponds to samples with acceptable key consistency but severe harmonic-progression errors (D3 $\geq$ 4 and D2 $\leq$ 2). A small deterministic jitter is applied to overlapping points for visualization only; all ratings remain integer-valued.}
\label{fig:d3d2}
\end{figure}

\subsection{Feature--Rating Correlations}

Table~\ref{tab:correlations} shows the nine prespecified Spearman correlations. The $p$-values are exploratory and uncorrected. A Bonferroni reference threshold for nine comparisons is 0.0056; no feature reaches it.

\begin{table}[t]
\centering
\caption{Spearman correlations between features and expert ratings ($n = 30$; $p$-values uncorrected). In the Expected Dir. column, $\checkmark$ indicates that the observed sign matches the hypothesized direction and $\times$ indicates that it does not.}
\label{tab:correlations}
\begin{tabular}{llrrll}
\toprule
Dim. & Feature & $\rho$ & $p$ & Sig. & Expected Dir. \\
\midrule
D1 & LLR      & $-0.429$ & 0.018 & nominal * & negative $\checkmark$ \\
D1 & PS       & $-0.013$ & 0.947 & n.s.      & negative $\checkmark$ \\
D2 & IKNR     & $+0.136$ & 0.475 & n.s.      & positive $\checkmark$ \\
D3 & KC       & $+0.300$ & 0.107 & n.s.      & positive $\checkmark$ \\
D3 & IKNR     & $+0.263$ & 0.161 & n.s.      & positive $\checkmark$ \\
D3 & KCR      & $-0.296$ & 0.112 & n.s.      & negative $\checkmark$ \\
D5 & LUFS\_dev & $+0.294$ & 0.115 & n.s.      & negative $\times$ \\
D5 & LRA      & $-0.039$ & 0.840 & n.s.      & positive $\times$ \\
D5 & SC       & $+0.040$ & 0.832 & n.s.      & positive $\checkmark$ \\
\bottomrule
\end{tabular}
\end{table}

\paragraph{D1: Large melodic leaps show a nominal association with expert ratings.}
LLR correlated negatively with D1 ($\rho = -0.429$, uncorrected $p = 0.018$): covers with more large jumps in the transcribed vocal line tended to receive lower melodic ratings. The association does not survive the nine-test multiplicity reference and is therefore treated as a feature-level lead, not validated construct evidence.

PS did not correlate with D1 despite also measuring interval behavior. LLR counts specific large-interval violations, whereas PS measures general interval variability, which can be high in stylistically varied but acceptable melodies. The ratings were more closely associated with salient implausible leaps than with general interval variance.

\paragraph{D2/D3: Key-related features are weak proxies.}
D2 IKNR was small ($\rho = 0.136$, $p = 0.475$). KC, IKNR, and KCR for D3 moved in the expected directions, but none reached uncorrected $p < 0.05$. At $n = 30$, these values do not establish reliable automatic indicators, and IKNR in particular cannot measure chord function.

\paragraph{D5: Low-level acoustic features do not explain arrangement quality.}
LUFS\_dev, LRA, and SC did not meaningfully correlate with D5 ratings, and LUFS\_dev moved opposite to the expected direction. Loudness normalization mainly reflects post-processing: a cover can sit near the target loudness while retaining poor instrumentation, weak texture, or unnatural timbre. Arrangement evaluation therefore needs musical and timbral relations that these summary statistics omit.

\subsection{Rule-Based Diagnostic Pilot}

Table~\ref{tab:rules} reports the accuracy comparison against the fixed majority-class baseline. The three-class rules did not beat the baseline on any dimension. Binary severe-error detection was directionally better for D2 and D3, but the paired intervals remained wide at $n = 30$.

\begin{table}[t]
\centering
\caption{Rule-based diagnostic accuracy versus fixed majority-class baseline. $\Delta$ is rule minus baseline; confidence intervals and $p$-values come from 10,000 paired generated-sample-level bootstrap draws.}
\label{tab:rules}
\begin{tabular}{llrrrl@{\hspace{6pt}}r}
\toprule
Dim. & Setting & Rule Acc. & Base Acc. & $\Delta$ & 95\% CI & $p$ \\
\midrule
D1 & 3-class & 0.433 & 0.467 & $-0.033$ & $[-0.233,\, 0.167]$ & 0.605 \\
D2 & 3-class & 0.433 & 0.533 & $-0.100$ & $[-0.333,\, 0.167]$ & 0.368 \\
D3 & 3-class & 0.433 & 0.633 & $-0.200$ & $[-0.433,\, 0.067]$ & 0.089 \\
D5 & 3-class & 0.300 & 0.467 & $-0.167$ & $[-0.400,\, 0.067]$ & 0.128 \\
D1 & Binary  & 0.600 & 0.700 & $-0.100$ & $[-0.233,\, 0.033]$ & 0.086 \\
D2 & Binary  & 0.567 & 0.533 & $+0.033$ & $[-0.267,\, 0.300]$ & 0.733 \\
D3 & Binary  & 0.767 & 0.700 & $+0.067$ & $[-0.100,\, 0.233]$ & 0.303 \\
D5 & Binary  & 0.433 & 0.533 & $-0.100$ & $[-0.333,\, 0.133]$ & 0.327 \\
\bottomrule
\end{tabular}
\end{table}

\begin{figure}[t]
\centering
\includegraphics[width=0.85\textwidth]{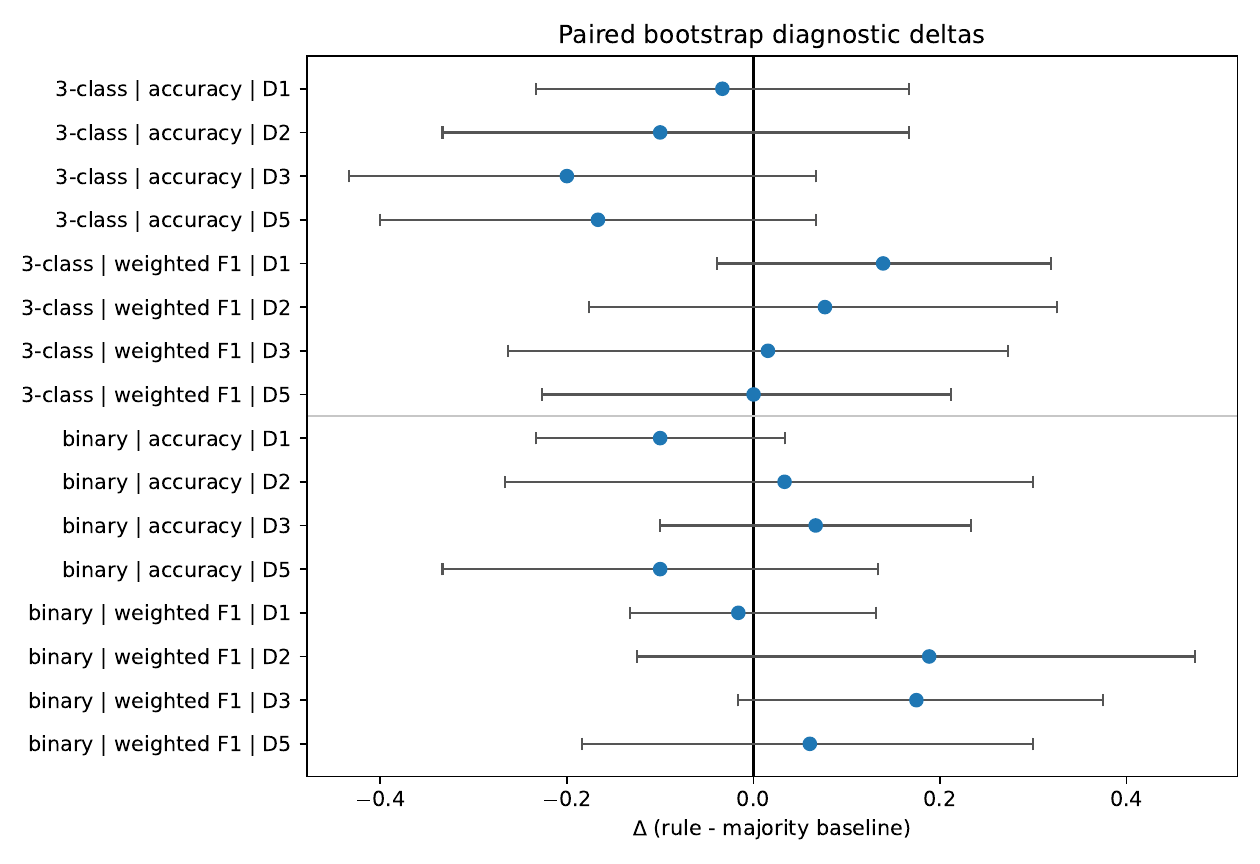}
\caption{Paired-bootstrap rule-minus-baseline differences. Points show observed differences in accuracy or weighted $F_1$, horizontal lines show 95\% paired-bootstrap intervals, and the vertical line marks zero. The plot contains the 16 dimension-level comparisons (4 dimensions $\times$ 2 settings $\times$ 2 metrics); aggregate overall summaries are excluded.}
\label{fig:bootstrap}
\end{figure}

Across the 16 dimension-level comparisons, every 95\% interval crossed zero and no uncorrected two-sided $p$-value was below 0.05. The largest positive trends occurred for binary weighted $F_1$ on D2 ($\Delta = 0.188$, 95\% CI $[-0.125, 0.473]$, $p = 0.232$) and D3 ($\Delta = 0.175$, 95\% CI $[-0.017, 0.375]$, $p = 0.072$), but neither supports a confirmed improvement. In the descriptive pooled summary, three-class accuracy was lower for the rule system than for the baseline (0.400 vs.\ 0.525; $\Delta = -0.125$, 95\% CI $[-0.233, -0.017]$, nominal $p = 0.023$). This pooled value is not treated as an independent dimension-level test. The 16 dimension-level $p$-values are reported without multiplicity adjustment for transparency; under a Bonferroni reference threshold of $0.05/16 = 0.0031$, no comparison would survive correction. This statement is included to avoid selective interpretation of nominal trends.

Global percentile thresholds discard musical context. IKNR cannot distinguish functional harmony from scale membership; key-confidence statistics cannot determine whether a local modulation is intentional; loudness and spectral summaries do not represent instrumentation, texture, or arrangement completeness. Even the nominal D1--LLR association did not produce reliable threshold classification against the baseline. The binary D3 result leaves open a possible use for coarse severe-error flags, but the present evidence does not support naive thresholding as a standalone evaluator.

\subsection{Illustrative Cases}

\paragraph{Case A: Complete success.}
MiniMax, Song~3, scored 5/5/5/5/5. The melody contour was natural and accurate, harmonic progressions were coherent, key was stable, the output matched the style prompt, and the arrangement showed appropriate timbre, groove, balance, and completeness.

\paragraph{Case B: Complete failure.}
ACE-Step+MCL, Song~1, scored 1/1/1/1/1. The output was out of tune, harmonically incoherent, and arrangement-deficient, with only drums clearly audible and severe timbral distortion in the remaining instrumentation.

\paragraph{Case C: Asymmetric failure.}
ACE-Step, Song~3, scored 4/2/5/5/2. The melody was plausible and the key was stable, but the accompaniment used B$\flat$ at 0:22--0:24 and 0:49--0:51 where the source progression uses G as the dominant resolving toward C~minor. B$\flat$ is diatonic to C~minor, but in these phrases it replaced the source song's functional dominant resolution. The arrangement was also sparse and poorly balanced. A single global score would obscure this profile: the output was not uniformly bad, but its harmonic and arrangement failures would be highly relevant for model debugging.

\section{Discussion}

\subsection{Diagnostic Evaluation Reveals Non-Uniform Failure}

Failures were not evenly distributed across musical dimensions. Harmonic progression and arrangement quality produced the most severe ratings, while key consistency was relatively preserved. The evaluated systems more often maintained a broad tonal center than a functional chord progression or complete arrangement.

\subsection{Key Consistency Is Not Harmonic Correctness}

Key consistency is not a proxy for harmonic correctness. D3 measures tonal stability, whereas D2 asks whether the chord sequence functions coherently within that center. The six D3-high/D2-low covers show that a cover can remain broadly ``in key'' while its harmonies fail to support the melody or phrase structure.

Pitch-class distributions and tonal-center cues therefore provide incomplete training targets. A system may remain in C~minor without learning when a dominant, subdominant, or chromatic alteration is contextually appropriate. Symbolic chord conditioning and phrase-level harmonic planning address that missing relation more directly.

\subsection{Automatic Features Capture Only Part of Expert Judgment}

The feature correlations and rule-based pilot show the limits of the present representation. LLR remains interpretable because unusually large vocal leaps are concrete and inspectable, but its association with D1 did not produce a threshold classifier that reliably beat the majority baseline. Correlation with an expert score is therefore not sufficient evidence that a feature can serve as an automatic decision rule.

The negative result is stronger for D2 and D5. IKNR describes scale membership, not chord function, while loudness, dynamic range, and spectral contrast describe signal distributions rather than whether a bass line supports the harmony, an instrument timbre matches the intended style, or an arrangement is structurally complete. Thresholding these summaries cannot recover the missing musical relations. Their proper role in the current framework is supporting evidence for expert diagnosis, not replacement of expert judgment.

\subsection{Implications for Evaluation and Product Use}

Global ratings should be supplemented with diagnostic dimensions. A moderate average can hide a consequential failure pattern, such as preserved melody paired with unusable harmony.

For product evaluation, the dimensions also separate recoverable defects from failures that require regeneration. Minor loudness or balance problems may respond to post-processing; severe harmonic errors usually require a new generation or a model-level correction.

\subsection{Future Work: Learned Representations Rather Than Threshold Tuning}

The pilot rules were intentionally simple and interpretable. Their failure to outperform the majority baseline reliably rules out further cutpoint tuning as a useful direction for this dataset. With only 30 covers, threshold optimization, ad hoc feature selection, or a small supervised classifier would mostly fit the benchmark's noise.

Future models need representations that preserve musical context: temporal embeddings relating melody to accompaniment, chord- and phrase-aware representations for harmonic function, and audio-text or style embeddings for D4. Training and testing them will require a larger multi-annotator corpus with held-out source songs. The current benchmark defines the failure modes but is too small for that model-development stage.

\section{Limitations}

\paragraph{Single annotator.}
All expert ratings were assigned by the first author. Although the first author has absolute pitch, formal music-theory training, and professional AI-music evaluation experience, inter-rater reliability cannot be assessed in the current version. The six-sample repeat annotation was a procedural consistency check, not a substitute for independent raters. The time-stamped annotation protocol improves transparency, but future work should include multiple annotators and agreement analysis.

\paragraph{Small sample size.}
The benchmark contains 30 samples from 5 songs and 6 systems. This is sufficient for an exploratory diagnostic study, but it limits statistical power. Feature--rating correlations should therefore be interpreted as preliminary evidence rather than definitive metric validation.

\paragraph{Source-song clustering.}
The 30 generated covers are grouped within 5 source songs, so outputs derived from the same source may share difficulty, structure, or genre effects. The paired bootstrap resamples generated-cover filenames, not source-song clusters. With only five clusters, cluster-bootstrap inference would itself be unstable; larger studies should sample more source songs and treat source identity as the primary resampling unit.

\paragraph{Transcription artifacts.}
Basic Pitch transcription may introduce octave errors, false notes, or artifacts from instrumental bleed. These issues can affect pitch-derived features, especially PR and PS. The study therefore treats computational features as auxiliary evidence rather than as ground truth.

\paragraph{No direct automatic D2 or D4 metric.}
Harmonic progression validity and style consistency are not represented by direct automatic measures. The pilot uses IKNR only as a deliberately weak D2 tonal proxy, not as a measure of chord function. This is a deliberate design choice: unreliable chord recognition or genre matching could produce misleading evidence. Future work should investigate more robust symbolic and learned representations for these dimensions.

\paragraph{Limited genre coverage.}
Five source songs cannot represent the diversity of popular music, harmonic language, production style, or cultural listening expectations. Broader claims require a larger and more varied source set.

\section{Conclusion}

This paper presented a five-dimensional diagnostic evaluation framework for AI-generated cover songs. Applied to 30 generated covers from 5 songs and 6 systems, the framework shows that the most frequent severe failures are not simply surface-level audio defects, but harmonic-progression and arrangement failures. Key consistency is comparatively better preserved.

The automatic analyses produced a narrower result than the feature correlations alone might imply. Large-leap ratio had a nominal relationship with melodic ratings, yet neither it nor the other features yielded rule-based decisions that reliably beat the majority baseline. Key summaries omit harmonic function, and loudness or spectral summaries omit arrangement structure.

AI cover evaluation must distinguish tonal stability from harmonic correctness and signal quality from musical coherence. A cover can remain in key while making locally wrong harmonic decisions. Diagnostic annotation exposes that difference now; dependable automation will require representations that model relations across melody, harmony, arrangement, and style.

The annotation schema, extracted features, and analysis pipeline are available at: \url{https://github.com/TiaaL/songecho-cover-metrics}.

\bibliographystyle{plainnat}
\bibliography{references}

\clearpage
\appendix

\section{Rule-Based Threshold Audit}
\label{app:thresholds}

Section~5.5 describes the percentile-rule pilot's threshold estimation. Because small-sample overlap can reverse the raw candidate order, every metric pair was reordered by its known direction before assignment. Table~\ref{tab:threshold_audit} records the metric properties and direction-enforcement policy. Exact per-fold cutpoint values are archived in the released JSON file.

\begin{table}[h]
\centering
\caption{Threshold audit: metric directions and reordering policy. ``Raw candidates'' are the severity-1 median and severity-2 quartile estimated within each LOO fold. When small-sample overlap reverses the expected order, the pair is reordered by the known metric direction before assigning yellow/red decision regions.}
\label{tab:threshold_audit}
\small
\begin{tabular}{llllp{4.5cm}}
\toprule
Metric & Direction & Dim. & Assertion & Note \\
\midrule
LLR      & higher-is-worse & D1      & yellow $<$ red & Large values indicate implausible leaps \\
PS       & higher-is-worse & D1      & yellow $<$ red & Large values indicate irregular intervals \\
IKNR     & lower-is-worse  & D2, D3  & yellow $>$ red & Low values indicate off-key notes \\
KC       & lower-is-worse  & D3      & yellow $>$ red & Low values indicate weak tonal center \\
KCR      & higher-is-worse & D3      & yellow $<$ red & High values indicate key instability \\
LUFS\_dev & higher-is-worse & D5      & yellow $<$ red & Large deviation from $-12$~LUFS reference \\
LRA      & lower-is-worse  & D5      & yellow $>$ red & Low values indicate compressed dynamics \\
SC       & lower-is-worse  & D5      & yellow $>$ red & Low values indicate flat spectral profile \\
\bottomrule
\end{tabular}
\end{table}

PR is excluded from the rule pilot because it is reported for reference only and is not used as a diagnostic metric. The per-fold JSON archive retains both raw candidates and direction-ordered cutpoints for reproducibility.

\section{Inter-Dimension Correlation Matrix}
\label{app:correlations}

Section~6.3 reports that all ten pairwise inter-dimension correlations had uncorrected $p < 0.01$. Table~\ref{tab:corr_matrix} provides the complete $\rho$ and $p$ values.

\begin{table}[h]
\centering
\caption{Pairwise Spearman rank correlations among the five diagnostic dimensions ($n = 30$). All correlations are positive and significant at uncorrected $p < 0.01$. These values are exploratory and are not adjusted for multiplicity.}
\label{tab:corr_matrix}
\begin{tabular}{lccccc}
\toprule
 & D1 & D2 & D3 & D4 & D5 \\
\midrule
D1 & --- & .728 & .727 & .507 & .761 \\
D2 &     & --- & .706 & .631 & .785 \\
D3 &     &     & --- & .688 & .799 \\
D4 &     &     &     & --- & .771 \\
D5 &     &     &     &     & --- \\
\bottomrule
\end{tabular}

\vspace{6pt}

\begin{tabular}{lcc}
\toprule
Pair & $\rho$ & $p$ \\
\midrule
D1--D2 & 0.728 & $< 0.0001$ \\
D1--D3 & 0.727 & $< 0.0001$ \\
D1--D4 & 0.507 & 0.0042 \\
D1--D5 & 0.761 & $< 0.0001$ \\
D2--D3 & 0.706 & $< 0.0001$ \\
D2--D4 & 0.631 & 0.0002 \\
D2--D5 & 0.785 & $< 0.0001$ \\
D3--D4 & 0.688 & $< 0.0001$ \\
D3--D5 & 0.799 & $< 0.0001$ \\
D4--D5 & 0.771 & $< 0.0001$ \\
\bottomrule
\end{tabular}
\end{table}

The weakest association is D1--D4 ($\rho = 0.507$, $p = 0.0042$); the strongest is D3--D5 ($\rho = 0.799$, $p < 0.0001$). The consistently positive structure indicates that model failures tend to co-occur, but the D3-high/D2-low asymmetric cases (Section~6.3) show that co-occurrence does not make the dimensions interchangeable.

\end{document}